\documentclass[showpacs,aps,prl,reprint,superscriptaddress,nofootinbib]{revtex4-1}
\usepackage[pdftex]{graphicx,color}

\usepackage{amsmath,amssymb,bm,color,longtable,mathrsfs}
\usepackage{ulem}
\usepackage[utf8]{inputenc}

\begin{document}
\begin{flushright}
DESY 19-213
\end{flushright}

\title{
Renormalization group on a triad network
}

\author{Daisuke~Kadoh}
\email[]{{\tt kadoh@keio.jp}}
\affiliation{Physics Division, National Center for Theoretical Sciences, National Tsing-Hua University, Hsinchu, 30013, Taiwan\\
Research and Educational Center for Natural Sciences, Keio University, Yokohama 223-8521, Japan}

\author{Katsumasa~Nakayama}
\email[]{{\tt katsumasa.nakayama@desy.de}}
\affiliation{NIC, DESY Zeuthen, Platanenallee 6, 15738 Zeuthen, Germany}

\date{\today}

\begin{abstract}
We propose a new renormalization scheme of tensor networks made only of third order tensors. 
The isometry used for coarse-graining the network can be prepared at an $O(D^6)$ computational cost
in any $d$ dimension ($d \ge 2$), where $D$ is the truncated bond dimension of tensors. 
Although it is reduced to  $O(D^5)$ if a randomized singular value decomposition is employed,  
the total cost is $O(D^{d+3})$ 
because the contraction part for creating a renormalized tensor with isometries has $D^{d+3}$ 
multiplications. 
We test our method in  three dimensional Ising model and find 
that the numerical results are obtained for large $D$s with reasonable errors.
\end{abstract}


\maketitle

\section{Introduction}
The tensor network is a promising approach to investigate statistical 
systems to which the Monte Carlo method is not easily accessible.
Since this approach is free from the sign problem, it is expected to be an 
essential tool to study finite density QCD, the theta vacuum, chiral gauge theories 
and supersymmetric models, and the real-time dynamics of field theories.
Since the tensor renormalization group (TRG) was proposed 
by Levin and Nave \cite{Levin:2006jai},
it has been improved 
\cite{2009PhRvB..80o5131G, Evenbly2015, RandTRG,Nakamura:2018enp, Adachi:2019paf, Lan:2019stu},
and the TRG and some related methods achieve success in studying two-dimensional lattice field theories
\cite{
Verstraete:2004zza,
Verstraete:2004cf,
Shimizu:2014uva, Shimizu:2014fsa, Takeda:2014vwa, 
Pichler:2015yqa,
Kawauchi:2016xng, Banuls:2016gid, Sakai:2017jwp, Banuls:2017ena, Shimizu:2017onf,  Kadoh:2018hqq, 
Kuramashi:2018mmi, Kadoh:2018tis, Bruckmann:2018usp,
Kuramashi:2019cgs, Banuls:2019hzc, Meurice:2019ddf, 
Bazavov:2019qih,Banuls:2019bmf}.
However, since it is designed for two-dimensional networks, a new scheme with much less computational cost is 
needed in order to study theories on three and four dimensions.

The higher order TRG (HOTRG) \cite{HOTRG} is a typical example of renormalization schemes in higher dimensions.
An isometry is given by 
the higher order singular value decomposition (HOSVD),
and  a square lattice network is coarse-grained by taking the contraction between two tensors 
with the isometries.
This method is applicable to any $d$ dimension for $d \ge 2$, and 
the cost scales as $O(D^{4d - 1})$ where $D$ is the truncated bond dimension of tensors 
at renormalization steps.
Recently, another scheme named as an anisotropic TRG (ATRG) was proposed in \cite{Adachi:2019paf}.
Although the cost is reduced to $O(D^{2d + 1})$ with a randomized singular 
value decomposition (RSVD) (See \cite{rand_trunc,2016arXiv160802148E}) or other truncation method, 
it has larger errors than the HOTRG for fixed $D$. 
So further studies for making an algorithm with small costs and higher accuracy are needed.

In this paper, we propose a new tensor renormalization scheme 
by defining it on a tensor network made 
only of third order tensors. 
That network and renormalization groups on it, which are referred to as a triad network 
and Triad RGs in this paper, 
respectively, are not uniquely determined. 
We give an example of Triad RGs improving a HOTRG-type renormalization on a triad network. 
The computational cost is drastically reduced since building blocks of our method are third order tensors.
We find that the order of cost for making an isometry does not depend on the dimensionality, but on $O(D^6)$ 
in any $d$ dimension ($d\ge 2$). 
This is reduced to $O(D^5)$ with the RSVD. 
The main cost comes from the contraction of making a renormalized tensor with isometries, 
which scales as  $O(D^{d+3})$ with the RSVD. 
Then a naive memory usage is proportional to  $O(D^{d+2})$ if intermediate tensors 
of order $d+2$ are stored on a computer. 
We test our method in three-dimensional Ising model and find that numerical results 
are obtained for larger $D$s with reasonable errors.

This paper is organized as follows. We firstly present our algorithm in three dimensions.
Then we test it in three-dimensional Ising model and compare results to those obtained 
from the HOTRG and the ATRG. 
The RSVD used in our method, a review of HOTRG with HOSVD and 
an extension to $d$ dimension are presented 
in appendices.

\section{Algorithm}
%
%
%
%
%
%
We begin with presenting our algorithm in three dimensions 
starting from a square lattice network made of 
a sixth order tensor $T_{ijklmn} \in \mathbb{C}$,
where all indices $i,j,\cdots$ run from $1$ to $N$. 
An extension to any dimension is given in the appendix. 
Without loss of generality, $T$ may be expressed as 
a canonical polyadic decomposition (CPD):  
\begin{eqnarray}
T_{ijklmn} = \sum_{a = 1} ^r
W^{(1)}_{a i}
W^{(2)}_{a j}
W^{(3)}_{a k}
W^{(4)}_{a l}
W^{(5)}_{a m}
W^{(6)}_{a n}
\label{cpd}
\end{eqnarray}
where $r$ takes a minimum value in the canonical form,  which is called a tensor rank of $T$. 
A tensor is derived in this form for lattice models with nearest neighbor interactions in general.

The tensor network of $T$ is defined  
on a three dimensional squared lattice  $\Gamma=\{(n_1,n_2,n_3)| n_i \in \mathbb{Z}\}$:
\begin{eqnarray}
Z= {\rm Tr}  \prod_{n \in \Gamma} T_{x_n x^\prime_n y_n y^\prime_n z_n z^\prime_n}
\label{eq:z}
\end{eqnarray}
where $x_n$, $x_n^\prime$, $y_n$, $y_n^\prime$, $z_n$, $z_n^\prime$
are indices defined on links stemmed from the site $n$.
These indices satisfy 
$x^\prime_n=x_{n+\hat 1}$, $y^\prime_n=y_{n+\hat 2}$, $z^\prime_n=z_{n+\hat 3}$ 
where $\hat \mu$ stands for the unit vector of $\mu$ direction.
${\rm Tr} $ denotes the summation of all indices.
%
%
All tensors live on sites and  any link shared by two tensors is contracted.

The computational cost of tensor renormalization 
for $d$ dimensional square lattice network is high 
when $d$ increases in general because the contraction between two $2d$th order tensors 
takes a high cost. In order to reduce the cost, 
we formulate a renormalization group on a network made only 
of third order tensors.

\begin{figure}[htbp]
\begin{center}
  \includegraphics[width=8cm, angle=0]{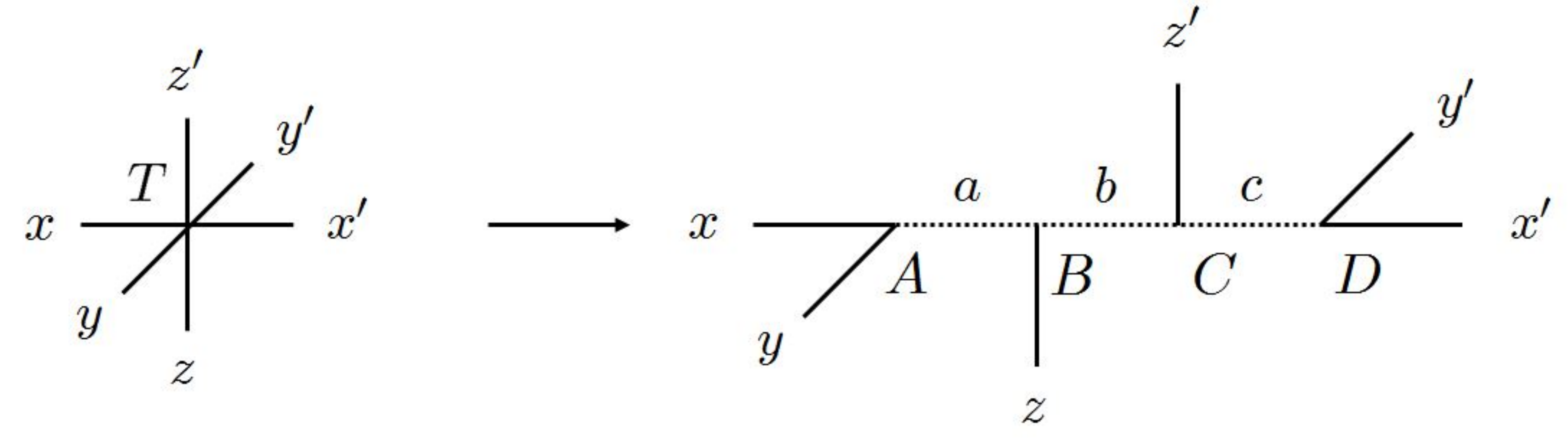}
 \caption{
Triad representation of $T$.
}
\label{fig:3rank}
\end{center}
\end{figure}

\begin{figure}[htbp]
\begin{center}
  \includegraphics[width=7cm, angle=0]{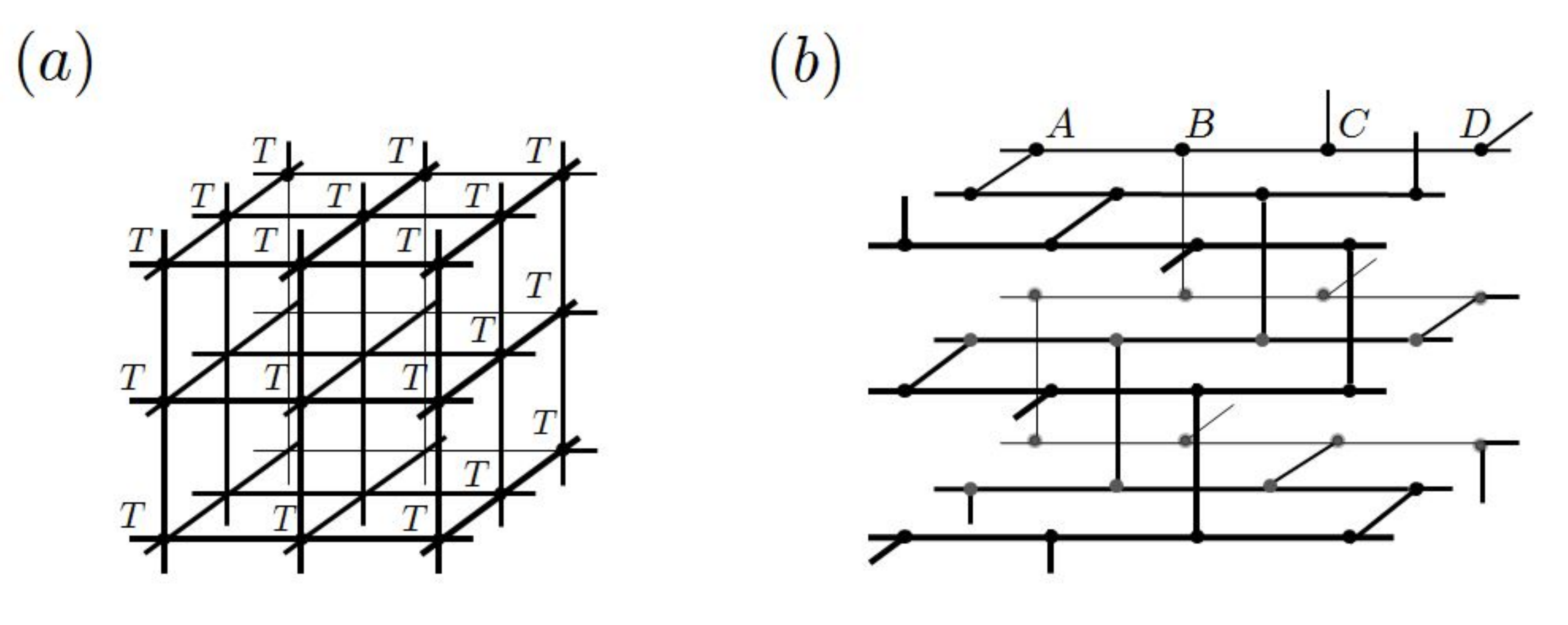}
 \caption{
Square lattice network $(a)$ and triad network $(b)$.
}
\label{fig:triad_network}
\end{center}
\end{figure}

It is possible to express $T$ as a product of four $3$rd order tensors $A,B,C,D$:
\begin{eqnarray}
&& T_{x x^\prime y y^\prime z z^\prime} = 
\sum_{a,b,c=1}^r
A_{x y a}  
B_{a z b} 
C_{b z^\prime c}
D_{c y^\prime  x^\prime},
\label{triad_unit}
\end{eqnarray}
where
\begin{eqnarray}
&& A_{xya}  \equiv W^{(1)}_{ax} W^{(2)}_{ay}, \\
&& B_{azb}  \equiv  \delta_{ab} W^{(3)}_{az}, \\
&& C_{bz c}  \equiv  \delta_{bc} W^{(4)}_{bz}, \\
&& D_{cyx}  \equiv W^{(5)}_{cy} W^{(6)}_{cx}.  
\end{eqnarray}
Eq.(\ref{triad_unit}) is referred to as 
a {\it triad} representation in this paper. 
Note that it is not unique. 
Fig.\ref{fig:3rank} shows a triad representation of $T$ shown in eq.(\ref{triad_unit}). 
Solid lines denote the external indices $x,y,z,\ldots$ and dotted lines denote the internal 
indices $a,b,c$.   
The tensor network $Z$ may be regarded as a network made only of third order tensors $A,B,C,D$ 
by replacing $T$ with a triad unit (\ref{triad_unit}) as shown in  Fig.\ref{fig:triad_network}. 
We also refer it to as a triad network. 

From now on, we assume that all internal and external indices, $a,b,c,...$ and $x,x^\prime,...$,
run from $1$ to $D$ for $D \ge r,N$. 
In renormalization steps, 
$D$ is a truncated bond dimension of triads $A,B,C,D$. 
Although a different size may be taken for internal indices 
to improve the accuracy of results, 
a common $D$ is employed and 
$\sum_{i=1}^D$ is denoted  as $\sum_i$ in the following for simplicity.

\begin{figure}[htbp]
\begin{center}
 \includegraphics[width=7cm, angle=0]{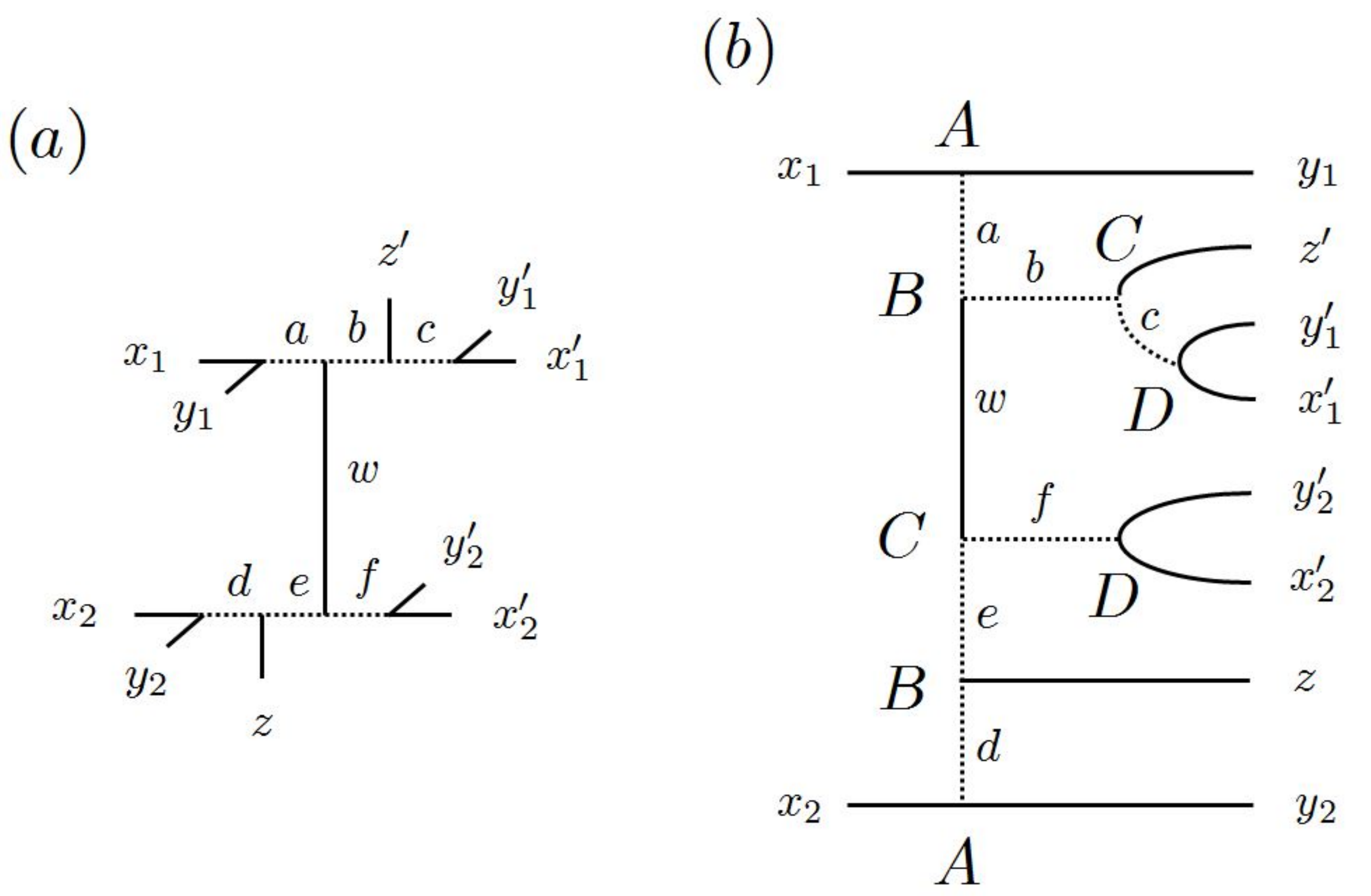}
 \caption{
$M$ in two triads. $(b)$ is another representation of $(a)$.   
}
\label{fig:M}
\end{center}
\end{figure}
We now consider a coarse-graining procedure of a Triad RG. 
A renormalization along the $z$ direction 
is carried out by combining two tensors as well as HOTRG:
\begin{eqnarray}
M_{X X^\prime Y Y^\prime  z z^\prime } 
=\sum_{w} T_{x_1 x_1^\prime y_1 y_1^\prime w z   } T_{x_2 x_2^\prime y_2 y_2^{\prime} z^\prime w}, 
\label{M}
\end{eqnarray}
where $X=x_1 \otimes x_2$ and $Y=y_1 \otimes y_2$. 
Fig.\ref{fig:M} shows $M$ in the triad representation. 
The HOSVD gives  
\begin{eqnarray}
&& M_{XX^\prime YY^\prime zz^\prime} = \sum_{I,J,K,L=1}^{D^2} \sum_{m,n=1}^D
S_{I JKLmn}  \nonumber \\
&& \qquad \times U^{L}_{XI} \, U^{R}_{ X^\prime J } \, V^{L}_{ YK} \, V^{R}_{Y^\prime L } 
\, W^{U}_{mz} \, W^{D}_{nz^\prime}
\label{HOSVD_M}
\end{eqnarray}
where $S$ is the core tensor of $M$ and $U, V,W$ are unitary matrices. 
See appendix for the detailed definition of HOSVD. 
$U$ and $V$ are used for coarse-graining the network but $W$ are not needed.

To evaluate $U^L$ associated with the $x$-axis, we interpret 
$M$ as a matrix $M^\prime_{X,\, X^\prime Y Y^\prime z z^\prime} (\equiv M_{X X^\prime Y Y^\prime z z^\prime}) $ 
with the row index $X$ and the column index $(X^\prime, Y,Y^\prime, z, z^\prime)$. 
Then $U^{L} $ is a unitary matrix that diagonalizes 
a hermitian matrix
\begin{eqnarray}
K \equiv M^\prime {M^\prime}^\dag
\label{Kmat}
\end{eqnarray}
as 
\begin{eqnarray}
K_{x_1x_2x_3x_4} = \sum_{I=1}^{D^2} U^L_{x_1x_2 I} \, \lambda^L_I \, U^{L\, \dag}_{I x_3x_4}, 
\label{diag_K}
\end{eqnarray}
where $\lambda_I$ are eigenvalues sorted in the descending order,
$\lambda^L_1 \ge \lambda^L_2 \ge \cdots \ge \lambda^L_{D^2} \ge 0$.
The cost of making $K$, which is ${\cal O}(D^{2d+2})$ in the HOTRG
as reviewed in the appendix,  
is drastically reduced on the triad network.

\begin{figure}[htbp]
\begin{center}
 \includegraphics[width=7cm, angle=0]{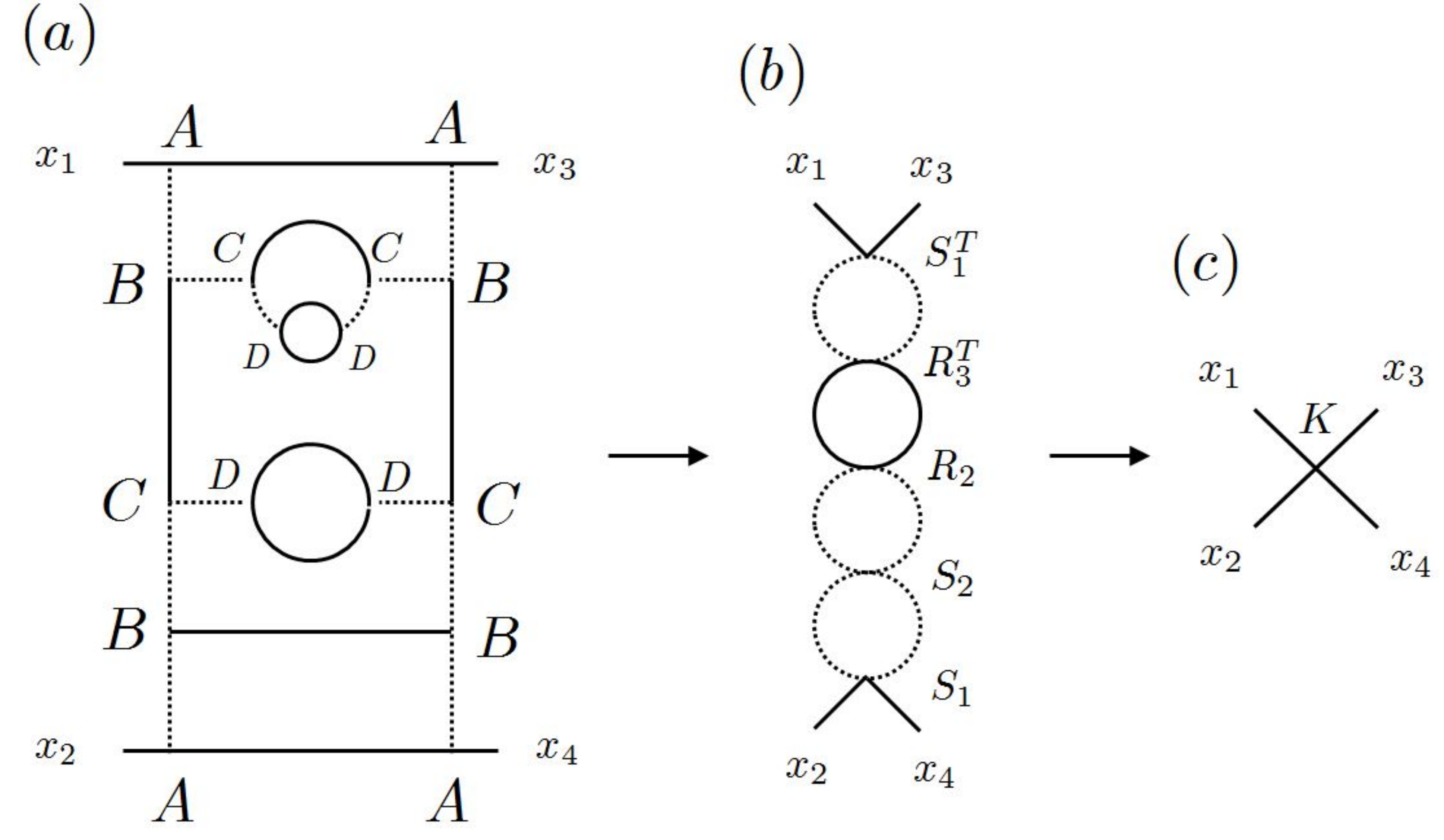}
 \caption{
How to make $K$ in the triad representation.  
}
\label{fig:3rank_proj}
\end{center}
\end{figure}
Fig.\ref{fig:3rank_proj} shows a methodology of making $K$ using the triad representation.
In Fig.\ref{fig:3rank_proj} (a), $K$ is represented as a product of $M$ (Fig.\ref{fig:M} (b)) 
and $M^\dag$ which is a mirror image of $M$. 
Fig.\ref{fig:3rank_proj} (b) is obtained by contracting inner lines connecting $M$ and $M^\dag$ 
from Fig.\ref{fig:3rank_proj} (a). 
When $K$ is expressed as a matrix $K^\prime_{x_2 x_4, x_1 x_3}(\equiv K_{x_1x_2x_3x_4})$ 
with row index $x_2,x_4$ and column index $x_1, x_3$, 
we have 
\begin{eqnarray}
 K^\prime = S_1  \cdot S_2  \cdot R_2  \cdot R_3^T  \cdot {S_1}^T,
 \label{K}
\end{eqnarray}
with hermitian matrices,  
\begin{eqnarray}
&& {(S_1)}_{x x^\prime, d d^\prime} = \sum_{y} A_{x y d} A^*_{x^\prime y d^\prime}, \nonumber \\
&& {(S_2)}_{dd^\prime, ee^\prime}  = \sum_{z} B_{d z e} B^*_{d^\prime z e^\prime}, \\
&& {(R_2)}_{ee^\prime, ww^\prime}  = \sum_{f,f^\prime}  C_{ewf} 
C^*_{e^\prime w^\prime f^\prime}  
\cdot 
 \sum_{x,y}  D_{f yx} D^*_{f^\prime yx},
\nonumber  \\
&& {(R_3)}_{aa^\prime, ww^\prime}  = \sum_{b,b^\prime} B_{awb}   
B^*_{a^\prime w^\prime b^\prime}  
\cdot \sum_{z} (R_2)_{bb^\prime,zz}.
\nonumber
\end{eqnarray}
Fig.\ref{fig:3rank_proj} (b) can be obtained at an ${\cal O}(D^5)$ cost without any approximation since 
$S_i$ and $R_i$ are computed at this cost. 
In Fig.\ref{fig:3rank_proj} (c), $K$ is obtained 
computing matrix products of 
eq.(\ref{K}), which takes an ${\cal O}(D^6)$ cost.    

We thus obtain $U^L$ at an $O(D^6)$ cost by diagonalizing 
the obtained $K$. 
The cost of making $K$ and $U^L$ can be reduced to $O(D^5)$ applying 
the RSVD to eq.(\ref{K}) (Fig.\ref{fig:3rank_proj} (b)), as shown in the appendix. 
The order of cost of making $U$  does not depend on the dimensionality 
in the triad representation.

The other unitary matrix $U^R$ can also be prepared in the similar manner.
We choose each one of $U^L$ and $U^R$ to improve the accuracy of results 
by comparing the remaining eigenvalues \cite{HOTRG}, 
\begin{eqnarray}
\epsilon_Q = \sum_{i>D} \lambda^Q_i,  \quad {\rm for} \ \ Q=L, R
\end{eqnarray}
where $\lambda^R$ are eigenvalues of $K^\prime$ 
with a different matrix representation 
$M_{X^\prime, X Y Y^\prime z z^\prime} (\equiv M_{X X^\prime Y Y^\prime z z^\prime}) $. 
$U=U^L$ for $\epsilon_L < \epsilon_R$ and $U=U^R$ for the others.

A renormalized tensor is defined by 
\begin{eqnarray}
\hspace{-2mm}
 T^R_{zz^\prime xx^\prime yy^\prime} \equiv \hspace{-3mm}
\sum_{X,Y,X^\prime,Y^\prime} \hspace{-2mm} 
U^\dag_{xX}  V^\dag_{yY}  
M_{XX^\prime YY^\prime zz^\prime}  U_{X^\prime x^\prime}  V_{ Y^\prime y^\prime} \ \
\label{TR}
\end{eqnarray}
where $U$ is $U^L$ (or $U^R$) and   $V$ is $V^L$ (or $V^R$) which are chosen from a comparison 
of $\epsilon_L$ and $\epsilon_R$. 
The combined indices $x,x^\prime,y,y^\prime$ run from $1$ to $D$ 
by truncating $D^2$ eigenvalues (functions) to $D$ largest ones.  

In the triad representation, we have
\begin{eqnarray}
&& T^R_{zz^\prime xx^\prime yy^\prime} = \sum_{a,b,e,f} 
{\cal D}_{zxya e } {\cal M}_{aebf} \, 
{\cal U}_{bf y^\prime z^\prime x^\prime}, 
\label{TR_triad}
\end{eqnarray}
where 
\begin{eqnarray}
&& {\cal U}_{ab yzx}= \sum_{c,p,q,p^\prime, q^\prime} 
C_{az c} 
D_{c qp} U_{pp^\prime x} 
D_{b q^\prime p^\prime} V_{q q^\prime y}, 
\label{U_def}
\\
&& {\cal M}_{abcd} = \sum_{w}  B_{awc} C_{bwd},\\
&& {\cal D}_{zxyab}= \sum_{d,p,q,p^\prime, q^\prime} B_{dzb}  
U^*_{pp^\prime x} A_{p q a} 
V^*_{q q^\prime y} A_{p^\prime q^\prime d}. 
\end{eqnarray}
Note that the cost for making ${\cal M}$ is $O(D^5)$ while
that for ${\cal U}$ and ${\cal D}$ is $O(D^6)$, which can be found 
by taking five contractions in the order of $(p,q^\prime)$ $\rightarrow$ $(p^\prime,q)$ $\rightarrow$ $c$ (or $d$) 
keeping intermediate fourth order tensors.

\begin{figure}[htbp]
\begin{center}
 \includegraphics[width=8cm]{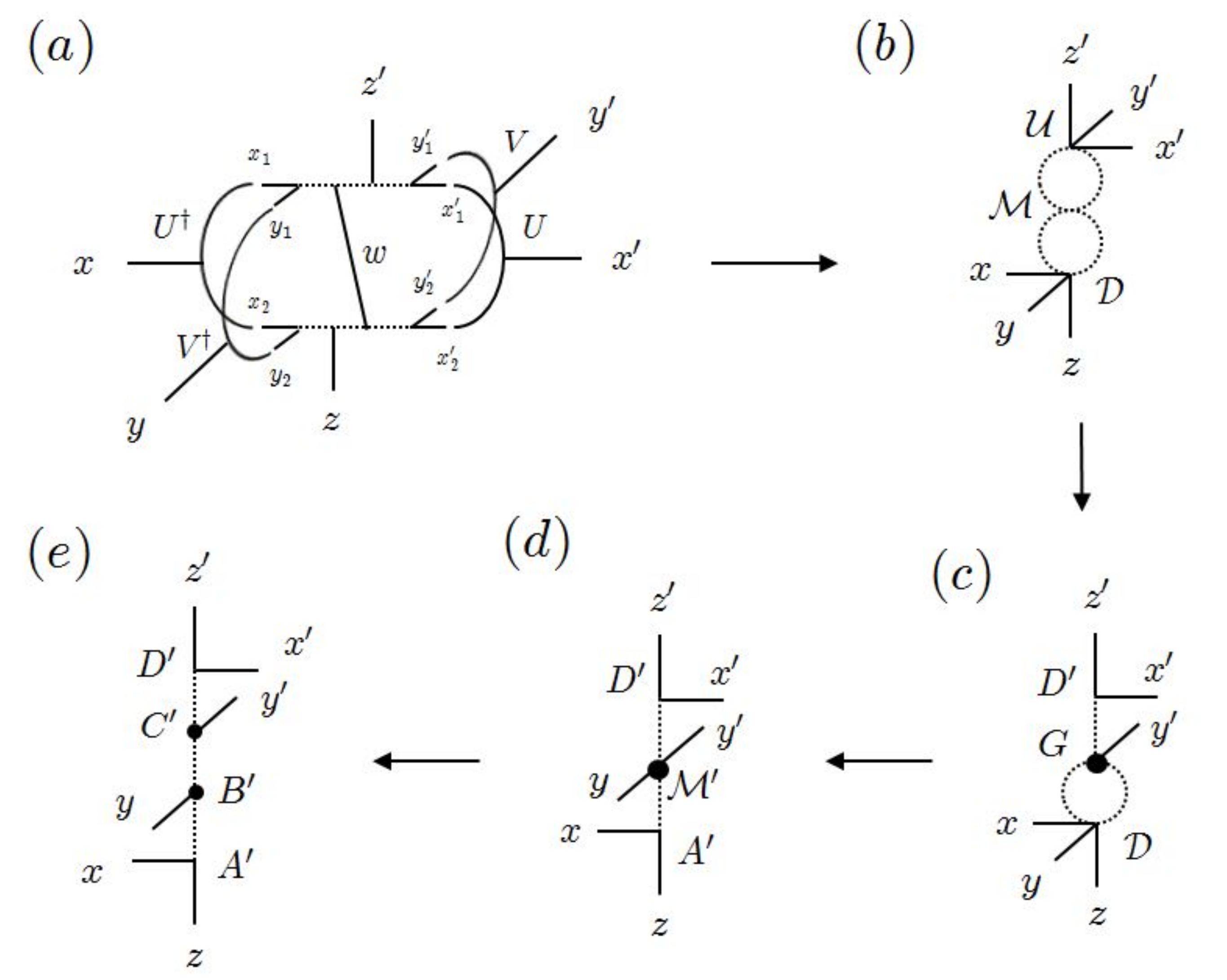}
 \caption{
Contraction of two triad units.
}
\label{fig:3rank_cont}
\end{center}
\end{figure}

Fig.\ref{fig:3rank_cont} shows how to create $T^R$.
Fig.\ref{fig:3rank_cont} (a) and (b)  show eq.(\ref{TR})  and eq.(\ref{TR_triad}), respectively.
Although a naive contraction between ${\cal M}$ and ${\cal U}$ in eq.(\ref{TR_triad}) takes $O(D^7)$, 
as presented in appendix, the RSVD provides an approximated decomposition  
at an $O(D^6)$ cost: 
\begin{eqnarray}
\sum_{c,d} {\cal M}_{abcd} \, {\cal U}_{cdyxz} \approx \sum_{g} G_{abyg} D^\prime_{gxz}
\label{step1}
\end{eqnarray}
where $D^\prime_{ijk}$ is a unitary matrix with the row $i$ and the column $j,k$ 
in terms of the full SVD and $G$ is a 4th order tensor in which the singular values are included. 
In Fig.\ref{fig:3rank_cont} (c), the black circle implies that $G$ contains 
the singular values. 

Then applying the RSVD to  remaining contractions in eq.(\ref{TR_triad}) 
in the similar manner,
we obtain another approximated decomposition at a cost of $O(D^6)$: 
\begin{eqnarray}
\sum_{a,b} {\cal D}_{zxya b} G_{ab y^\prime g}  \approx 
\sum_{e} A^\prime_{zxe} {\cal M}^\prime_{eyy^\prime g} 
\label{step2}
\end{eqnarray}
where $A^\prime_{ijk}$ is a unitary matrix with the row $i,j$ and the column $k$ 
in terms of the full SVD and ${\cal M}^\prime$ is a 
fourth order tensor in which the singular values are absorbed. 
Fig.\ref{fig:3rank_cont} (d) represents this decomposition.   
The SVD finally provides  
\begin{eqnarray}
{\cal M}^\prime_{eyy^\prime g} \approx \sum_{f} B^\prime_{eyf} C^\prime_{fy^\prime g} 
\label{step3}
\end{eqnarray}
at an $O(D^6)$ cost, as shown in Fig.\ref{fig:3rank_cont} (e). 
The RSVD reduces this cost to $O(D^5)$.

Plugging eqs.(\ref{step1})-(\ref{step3}) into  eq.(\ref{TR_triad}),  
we thus find that $T^{R}$ is approximately given 
by a renormalized triad unit: 
\begin{eqnarray}
&& T^R_{zz^\prime x x^\prime y y ^\prime} \approx \sum_{a,b,c} 
A^\prime_{z x a}  
B^\prime_{a y b} 
C^\prime_{b y^\prime c}
D^\prime_{c x^\prime  z^\prime}.
\label{renormalized_triad_unit}
\end{eqnarray}
An extra axis rotation is not needed since  $x,y,z$ of eq.(\ref{triad_unit}) is replaced by $z,x,y$ in 
(\ref{renormalized_triad_unit}).  
In three dimensions, the computational cost of the Triad RG method is $O(D^6)$, 
and a memory usage is naively $O(D^5)$, which comes from the fifth order tensors ${\cal U,D}$. 
Repeating this procedure again and again, triad networks can be coarse-grained 
at this cost keeping the triad representation.   

An extension to any $d$ dimension is presented in the appendix. 
Decompositions such as
eq.(\ref{step1}) take   $O(D^{d+3})$ as ${\cal U}$ and ${\cal D}$ are tensors of order $d+2$, 
which is the main cost of our Triad RG method. 
The memory usage is naively $O(D^{d+2})$
if ${\cal U}$ and ${\cal D}$ are stored in the memory.

We make a remark on improvements of the Triad RG method 
in the rest of this section. 
As shown in Fig.\ref{fig:3rank_cont} (c)-(e),  
the singular values denoted as the black circles are inherited 
to the next decomposition, as done in the ATRG. 
The best accuracy is achieved with this treatment 
as long as we tried. 
Although we have presented a $O(D^{d+3})$ procedure for the contraction part,
we can easily create another procedure with an $O(D^5)$ cost
neglecting the accuracy of results. 
For instance, decomposing ${\cal M}$ in eq.(\ref{TR_triad}) directly
or swapping indices of tensors with the RSVD can reduce
a naive cost.
We found that these additional SVDs do not work well
for the three dimensional Ising model. This could be because 
such SVDs are optimized only for local tensors and the accuracy 
for the whole two triad units  
is lost. 
However, since there are many variants in the contraction part of our method, 
further studies are needed to improve the Triad RG method.

\section{A Numerical test}
%
%
We test the Triad RG method in three dimensional Ising model on a periodic lattice 
with the volume $V = (32768)^3=(2^{15})^3$ 
at the critical temperature $T_c = 4.5115$ \cite{critem_1,HOTRG,Adachi:2019paf}.  
The Hamiltonian is given by 
\begin{eqnarray}
H=-\sum_{ \langle i,j \rangle} \sigma_i \sigma_j, 
\end{eqnarray} 
where $\langle i,j \rangle$ denotes all possible nearest neighbor pairs of lattice sites. 
The partition function $Z={\rm Tr}(e^{-\beta H})$ with the inverse temperature $\beta=1/T$ 
can be expressed as a triad network eq.(\ref{triad_unit})
with $N=r=2$ and $W^{(\mu)} \equiv W$ given by
\begin{eqnarray}
W=
\begin{pmatrix}
  \sqrt{\mathrm{cosh}(\beta )} & \sqrt{\mathrm{sinh}(\beta)} \\
  \sqrt{\mathrm{cosh}(\beta)} & -\sqrt{\mathrm{sinh}(\beta)}
\end{pmatrix}.
\end{eqnarray}
Note that initial 3rd order tensors $A,B,C,D$ are real and satisfy 
$A_{xya} = D_{ayx} $ and $ B_{azb} = C_{azb} $.   

The free energy is evaluated by $F=-\frac{1}{\beta} {\rm log} Z$. 
We take $C=4$ for the oversampling parameter of RSVD 
(See the appendix for details of RSVD).  
The numerical computation is carried out with 2.7 GHz Intel Core i7 
and a library {\it Eigen} for matrix decompositions, 
and each computation ends in a few hours. 

Numerical results are compared to those obtained from the HOTRG and the ATRG. 
The first version of ATRG is implemented by 
the RSVD with twice a larger bond dimension $2D$ 
only for the swapping step.
In the Triad RG, an $O(D^5)$ isometry is prepared with the RSVD. 
The computational cost of HOTRG, ATRG, and Triad RG methods 
are theoretically $O(D^{11})$, $O(D^7)$, and $O(D^6)$ in three dimensions, respectively.

\begin{figure}[htbp]
\begin{center}
 \includegraphics[width=5cm, angle=0]{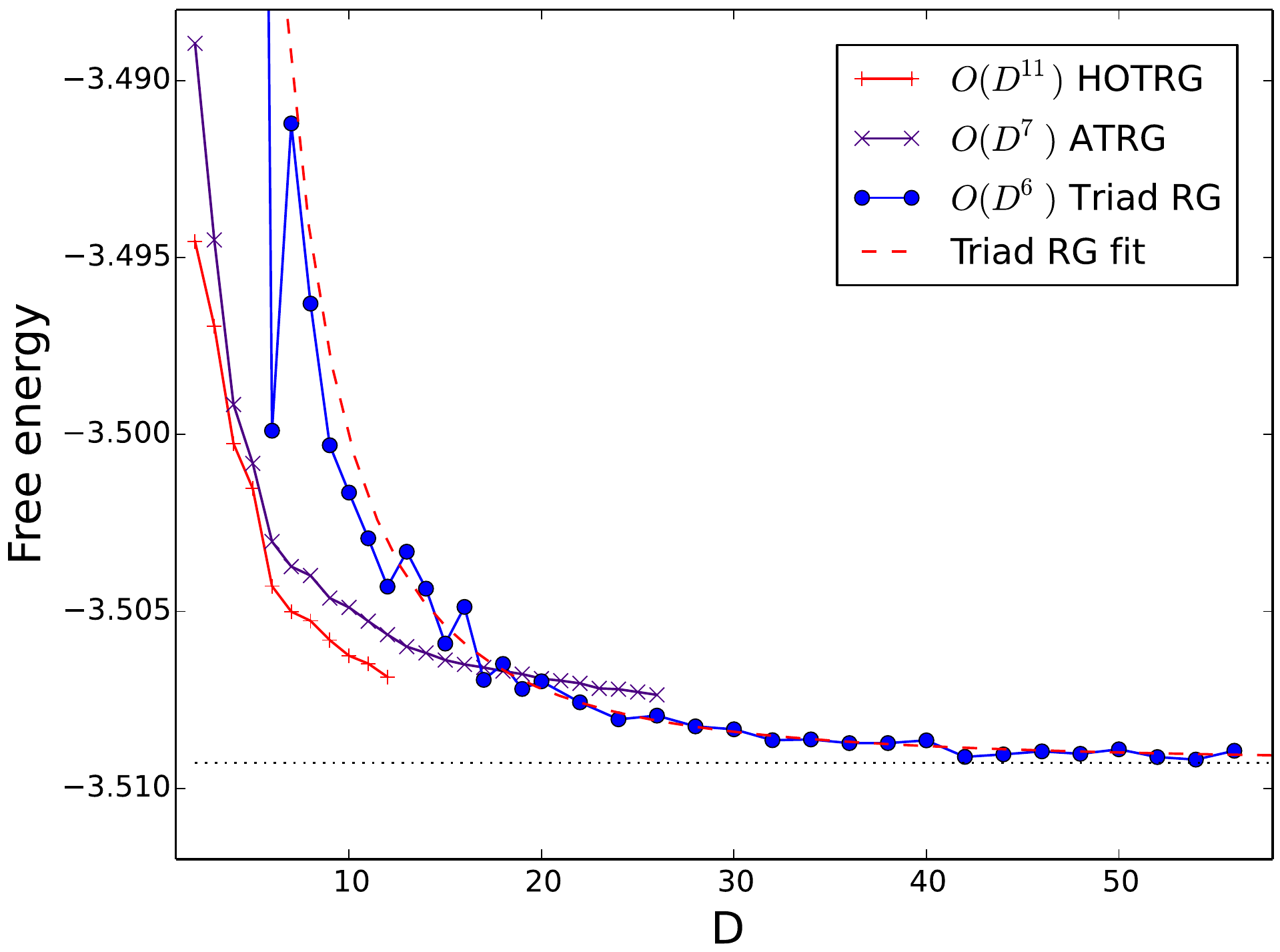}
 \caption{
$D$-dependence of free energy 
in 3d Ising model at $T_c$. 
}
\label{fig:Free_D}
\end{center}
\end{figure}
Fig.\ref{fig:Free_D} shows the $D$-dependence of free energy. 
The accessible $D$ is different among the three methods.
Three results decrease toward values around  $F = -3.51$
as $D$ increases. 
The Triad RG has well-controlled errors for larger $D$s and shows good convergence.

We extrapolate our result of the Triad RG to the large $D$ limit using a fit function $a + bD^{-c}$ 
with fitting variables $a,b,c$.
To obtain a precise fit result, we compute the free energy $n$ times with different random numbers of the RSVD 
($n=100$ for $D\leq 24$ and $n=4$ for  $D > 24$), 
and use an average value of $n$ trials with error estimated from the standard deviation for the fit. 
The result for $10 \leq D \leq 56$ is $a = -3.5093(2)$, which is shown 
as a dotted line in Fig.\ref{fig:Free_D}.
We confirm that this result is stable by changing fit range to $20 \leq D \leq 56$.

\begin{figure}[htbp]
\begin{center}
 \includegraphics[width=5cm, angle=0]{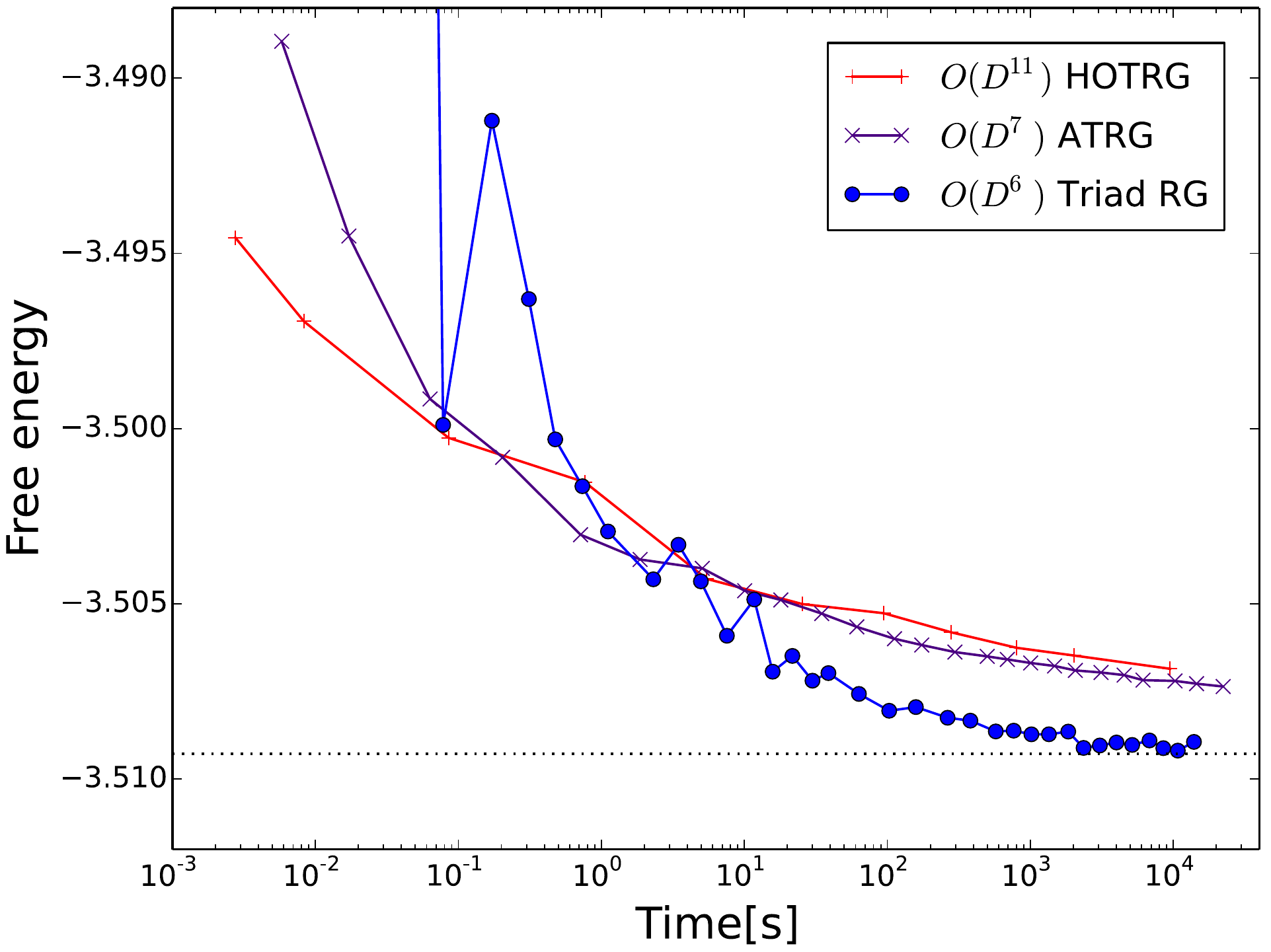}
 \caption{
Free energy as a function of computational time.
}
\label{fig:Free_err}
\end{center}
\end{figure}
Fig.\ref{fig:Free_err} shows the free energy against the computational time needed to compute the free energy once.  
This figure implies that the Triad RG converges faster than the other methods 
for the same computational time.
\begin{figure}[htbp]
\begin{center}
 \includegraphics[width=5cm, angle=0]{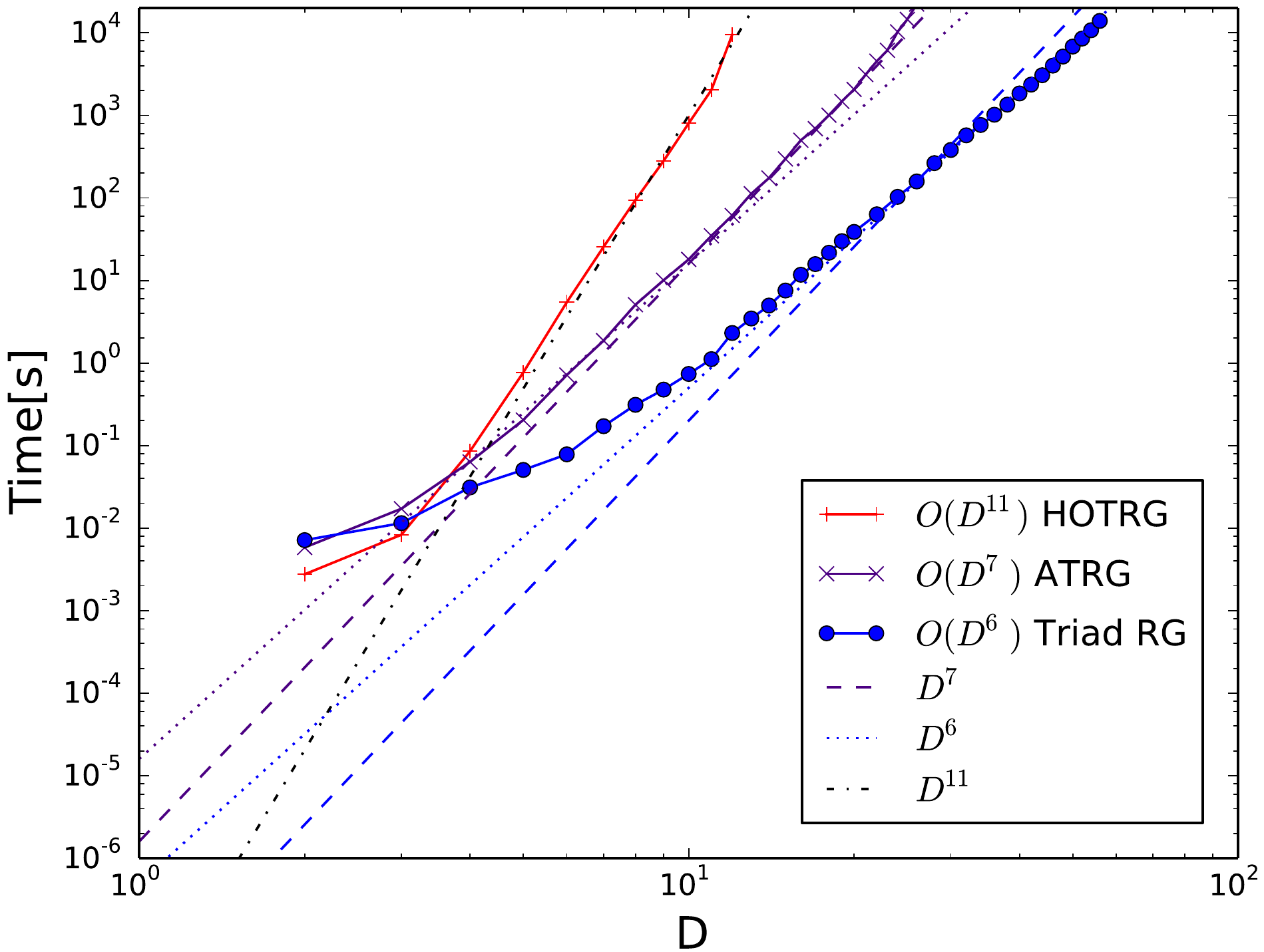}
 \caption{
Computational time against $D$.
}
\label{fig:time}
\end{center}
\end{figure}
In Fig.\ref{fig:time}, the computational time is shown as a function of $D$.  
Since the theoretical $D$-dependence is properly reproduced at a practical level,  
one can consider that the Triad RGs will open a door to studying 
a wide class of higher dimensional field theory with tensor networks.

\section*{Acknowledgments}
We would like to thank 
Shinichiro Akiyama, Pochung Chen, Karl Jansen, Ying-Jer Kao, Yoshinobu Kuramashi, C.-J. David Lin, Tomotoshi Nishino, Hideaki Oba, 
Ryo Sakai, Manuel Schneider, 
Amit Singh, Shinji Takeda, Hiroshi Ueda and Yusuke Yoshimura 
for their valuable comments.
D.K. is supported by JSPS KAKENHI Grant JP19K03853 and 
the National Center for Theoretical Sciences at National Tsing-Hua University. 
K.N. is supported partly by the Grant-in-Aid for Japan Society for the Promotion of Science Research Fellow (No. 18J11457).

\bibliography{SVD}

\appendix
\section{Randomized SVD and contraction of tensors}

The cost of decomposing (or contracting) tensors is reduced 
by a randomized SVD (RSVD) \cite{rand_trunc, 2016arXiv160802148E}. 
 In this appendix, we present technical details of RSVD 
and how it is used for making projectors $U$ in eq.(\ref{diag_K}) 
and for three decompositions eqs.(\ref{step1})-(\ref{step3}) in the Triad RG method.

Let $A$ be an $m \times n$ matrix with complex entries. 
The SVD provides a decomposition of $A$: 
\begin{eqnarray}
A=U\Sigma V^\dag, 
\end{eqnarray}
where $U$ and $V$ are $m \times m$  and $n \times n$ unitary matrices, respectively, and 
$\Sigma $ is a non-negative diagonal matrix containing singular values in the descending 
order, $\Sigma_{ij} =\sigma_i \delta_{ij}$ ($\sigma_1 \ge \sigma_2 \ge \ldots$).   
The cost of full SVD is $O(m n l)$ for $l={\rm min}\{m,n\}$.
However, we often need part of SVD such as $k$ largest singular values for $k \ll m,n$.   
Then the cost is reduced from $O(m n l)$ to $O(mnk)$ with the RSVD. 

For a  given $m\times n$ matrix $A$, $k$ largest singular values and singular vectors 
are approximately given by the following procedure (RSVD): \vspace{2mm}

1. Generate an $n\times p$ Gaussian random matrix $\Omega$. \vspace{2mm}

2. Construct an $m \times p$ matrix $Y_A = A \Omega$. \vspace{2mm}

3. Compute QR factorization of $Y_A$ as $Y_A=Q_A R_A$.\vspace{2mm}

4. Construct a small  $p \times n$ 
matrix $B_A \equiv Q^\dag_A A$.  \vspace{2mm}

5. Compute an SVD of $B_A$  as $B_A=\tilde U \tilde \Sigma \tilde V^\dag$. \vspace{3mm}

\noindent 
$p \, (\ge k)$ is a tunable parameter. 
Note that $R_A$ is an  $m \times p$ upper triangular matrix and 
$Q_A$ is an $m \times m$ unitary matrix (the first $m \times p$ part of $Q_A$ 
is needed in the fourth step).
We thus obtain a low rank approximation of $A$: 
\begin{eqnarray}
A \approx U^\prime \tilde \Sigma \tilde V^\dag
\label{approx_A}
\end{eqnarray} 
 where $U^\prime \equiv Q_A \tilde U$. 
$k$ largest singular values with singular vectors of $A$ are  
approximately given by taking $k$ part of eq.(\ref{approx_A}). 
In this sense, $p$ provides an oversampling of the index 
$k$ to improve the accuracy of results.

The costs of the second and fourth steps of RSVD are $O(mnp)$,
which are higher than the cost of QR decomposition at the third step $O(p^2m)$ 
and  the cost of the fifth step $O(p^2n)$. As long as we take 
\begin{eqnarray}
k \le p \ll m,n
\label{first_cond}
\end{eqnarray}
the main cost of RSVD, which is $O(pmn)$, is lower than the cost of full SVD. 
See \cite{rand_trunc,2016arXiv160802148E} 
for a formal discussion of error of RSVD.

The RSVD is also useful to evaluate a matrix product 
and to give its low-rank approximation. Let $E$ and $F$ be
$m \times \ell$ and $\ell \times n$ matrices, respectively.   
Although the cost of evaluating a matrix product $A=EF$ directly from $E$ and $F$ is $O(mn\ell)$,
a low-rank approximation of $A$ can be given at a lower cost 
using the same procedure of RSVD. 
At the second step,  we can construct $Y_A$ multiplying $\Omega$ by $E$ and $F$ in turn 
instead of $A$. These multiplications take $O((m+n) p\ell)$, 
which is smaller than $O(mn\ell)$ under eq.(\ref{first_cond}). 
Similarly, the fourth step takes the same cost. 
Once $Q_A$ and $B_A$ are obtained, 
without the fifth step,  
one can give a low rank approximation of $A=EF$ as
\begin{eqnarray}
A_{ij} \approx \sum_{a=1}^p (Q_A)_{ia} (B_A)_{aj}.
\label{mprod_approx} 
\end{eqnarray}
Note that  the contraction of rhs takes an $O(mnp)$ cost.
Although the matrix product $EF$ takes $O(mn\ell)$, 
we thus find that 
it is approximated by rhs of eq.(\ref{mprod_approx}) at a lower cost $O((mn+\ell n + \ell m)p)$ for 
\begin{eqnarray}
k \le p \ll \ell,m,n.
\label{second_cond}
\end{eqnarray}
Moreover, the fifth step provides the SVD of $EF$ at an $O((mn+\ell n + \ell m)p)$  cost. 

These tricks are used for tensors in the triad RG method. 
For the truncated bond dimension $D$, 
we basically take $m,n,\ell \ge D^{2}$ 
and $k=D$ and $p=Ck (= CD)$ where a fixed integer $C$ 
is referred to as an oversampling parameter in this paper. 
Since these parameters satisfy eqs.(\ref{first_cond}) and (\ref{second_cond}) for $D \gg 1$, 
the RSVD effectively works. 
To estimate the error, we compute a physical value $N$ times 
with different random numbers. The result with an error is evaluated 
from the central value and  the standard deviation from $N$ trials.

Let us consider a contraction between $4$th order tensor ${\cal M}$ 
and $5$th order tensor ${\cal U}$:  
\begin{eqnarray}
{\cal A}_{zx,aby} \equiv \sum_{c,d=1}^D {\cal M}_{abcd} \, {\cal U}_{cdyxz}.
\label{tensor_contraction}
\end{eqnarray} 
Note that $n=D^3,m,\ell=D^2, k=D$ and $p=CD$ in this case and eq.(\ref{second_cond}) is satisfied.  
 Although the cost of evaluating ${\cal A}$ 
from ${\cal M}$ and ${\cal U}$ is $O(D^{7})$,
a low rank approximation of ${\cal A}$  such as eq.(\ref{mprod_approx}) can be given 
at a cost of $O(D^6)$. Once ${\cal A}$  is approximately given, applying the RSVD to it 
like  eq.(\ref{approx_A}), we obtain
 \begin{eqnarray}
{\cal A}_{zx,aby}
\approx \sum_{g=1}^D G_{abyg} D_{g zx}
\label{tensor_approx}
\end{eqnarray}
with 
\begin{eqnarray}
&& G_{abyg} =  \tilde V^\dag_{g,aby} \tilde \sigma_g,\\
&& D_{g xz} = (Q_{\cal A} \tilde U)_{zx,g}, 
\end{eqnarray}
where $B_{\cal A}=\tilde U \tilde \Sigma  \tilde V^\dag$, 
at an $O(D^6)$ cost.
The singular values are included in $G$. 
This proves eq.(\ref{step1}).
Eqs.(\ref{step2}) and (\ref{step3}) are derived in a similar manner.

We apply the RSVD to
the $D^2 \times D^2$ matrix $K$ given in eq.(\ref{K}) in order to obtain  
$D$ singular vectors $U_{x_1x_2x}$ for $x_1,x_2,x=1,2\cdots,D$. 
Note that $m=n=\ell=D^2$, $k=D$ and $p=CD$ in this case and eq.(\ref{first_cond}) is satisfied.
We do not construct $K$ directly since it takes $O(D^6)$.  
Instead, we can evaluate $Y_K$ at an $O(D^5)$ cost 
multiplying $\Omega$ by $S_i$ and $R_i$ alternately where $\Omega$ is a 
$D^2 \times p$ Gaussian random matrix.  The fourth step takes the same cost. 
Once $Q_K$ and $B_K$ are given,  $K$ is approximately given at an $O(D^5)$ cost like eq.(\ref{mprod_approx}). 
Applying the RSVD to the constructed matrix $K_{x_1x_2,x_3x_4}$ again, 
we obtain a low rank approximation of $K$ at an $O(D^5)$ cost like eq.(\ref{approx_A}): 
\begin{eqnarray}
K_{x_1x_2x_3x_4} \approx \sum_{x=1}^{D} U^L_{x_1x_2 i} \, \tilde \lambda^L_i \, \tilde U^{L\, \dag}_{i x_3 x_4}, 
\end{eqnarray}
where $U^L \approx \tilde U^L$. 
This procedure can be easily extended to higher dimensions.   
We can obtain isometries at an $O(D^5)$ cost with the RSVD in any dimension.

\section{HOTRG}
\label{app:hotrg}
The HOTRG method \cite{HOTRG} with the HOSVD  is reviewed in this appendix. 
We will find out that the cost of $d$ dimensional HOTRG 
is proportional to  $D^{4d-1}$,
which comes from final contractions with isometries 
although the cost of making isometries scales as $D^{2d+2}$.  

To introduce the HOSVD, let us first define the inner product for tensors: 
\begin{eqnarray}
\langle {\cal T}, {\cal T}^\prime \rangle \equiv  \sum_{i_1,i_2,\cdots,i_n} 
{\cal T}^*_{i_1i_2\cdots i_n} {\cal T}^\prime_{i_1i_2\cdots i_n} 
\end{eqnarray}
where ${\cal T}$ and ${\cal T}^\prime$ are $n$th order tensors.  
The norm of ${\cal T}$ is defined as 
$\vert\vert {\cal T} \vert\vert = \sqrt{\langle {\cal T}, {\cal T} \rangle}$.
The HOSVD tells us that an $n$th order tensor  ${\cal T}_{i_1 i_2,\cdots i_n}$ 
($i_k=1,2,\cdots,N_k$)
may be expressed as 
\begin{eqnarray}
{\cal T}_{i_1i_2\cdots i_n} =   \sum_{j_1,j_2,\cdots,j_n} 
{\cal S}_{j_1j_2\cdots j_n} U^{(1)}_{i_1 j_1} U^{(2)}_{i_2 j_2}  \cdots U^{(n)}_{i_n j_n}
\end{eqnarray}
Here $U^{(m)}$ is an $N_m \times N_m$ unitary matrix and 
${\cal S}_{i_1i_2\cdots i_n}$ 
is  a core tensor of $T$, which satisfies 
 all-orthogonality:  
\begin{eqnarray}
\langle {\cal S}_{i_m=\alpha}, {\cal S}_{i_m=\beta} \rangle = 0,   \ \  {\rm for} \ \ \alpha \neq \beta,    
\end{eqnarray}
and an ordering property:
\begin{eqnarray}
\vert\vert {\cal S}_{i_m=1}\vert\vert \ge \vert\vert {\cal S}_{i_m=2}\vert\vert  
\ge \cdots \ge \vert\vert {\cal S}_{i_m=N_m} \vert\vert \ge 0, 
\end{eqnarray}
where $ {\cal S}_{i_m=\alpha}$ is a sub tensor of order $n-1$ with the $m$-th index 
$i_m$ of $ {\cal S}$ is fixed to $\alpha$.

\begin{figure}[htbp]
\begin{center}
 \includegraphics[width=7cm]{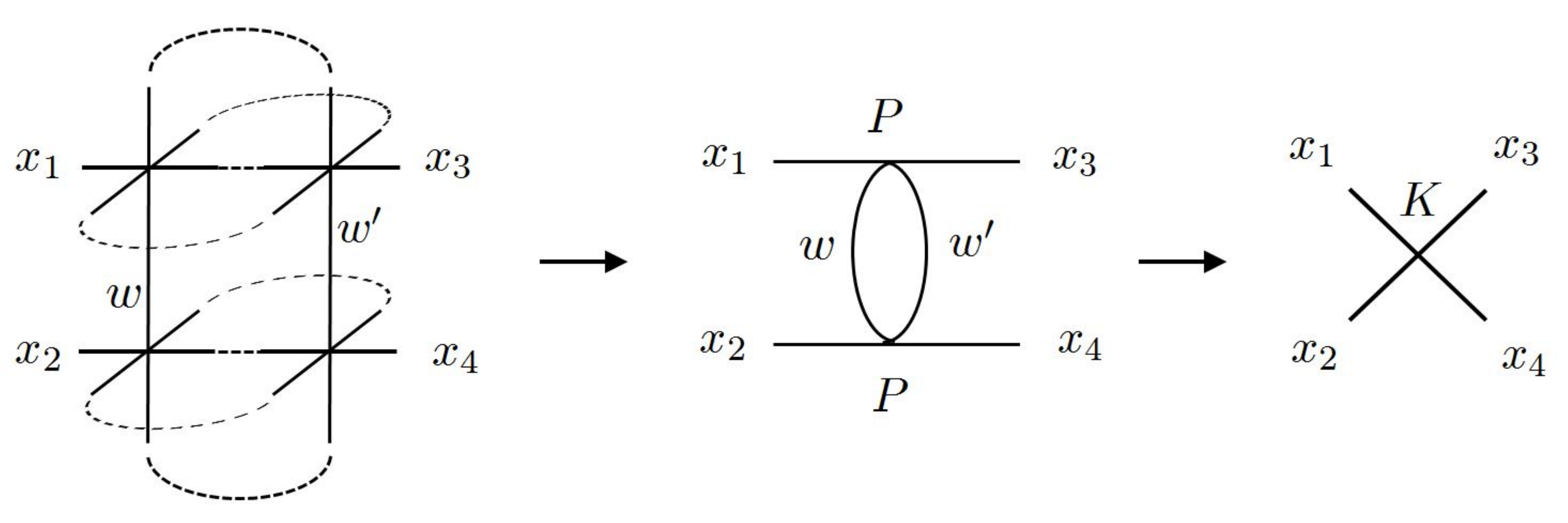}
 \caption{
A method of evaluating $K$ for a three dimensional square lattice network.
}
\label{fig:proj}
\end{center}
\end{figure}
The renormalization of the HOTRG is carried out for all axes alternately. 
$M$ in eq.(\ref{M}) is expressed as eq.(\ref{HOSVD_M})
in terms of the HOSVD. 
We evaluate 
$K$ defined by eq.(\ref{K})  to give isometries. 
Fig.\ref{fig:proj} shows how to create $K$ in three dimensions.
We can make a 4th order tensor $P$ by paying an $O(D^{2d+2})$ cost.   
The final step to make $K$ from two $P$s and the diagonalization of $K$ 
do not need a high cost. 
We find that the cost of making $U^L$ (or $U^R$) scales as $O(D^{2d+2})$.

A renormalized tensor is evaluated in a similar way to eq.(\ref{TR}).
In $d$ dimension, the number of isometries $U$ is $d-1$.

\begin{figure}[]
\begin{center}
 \includegraphics[width=7cm, angle=0]{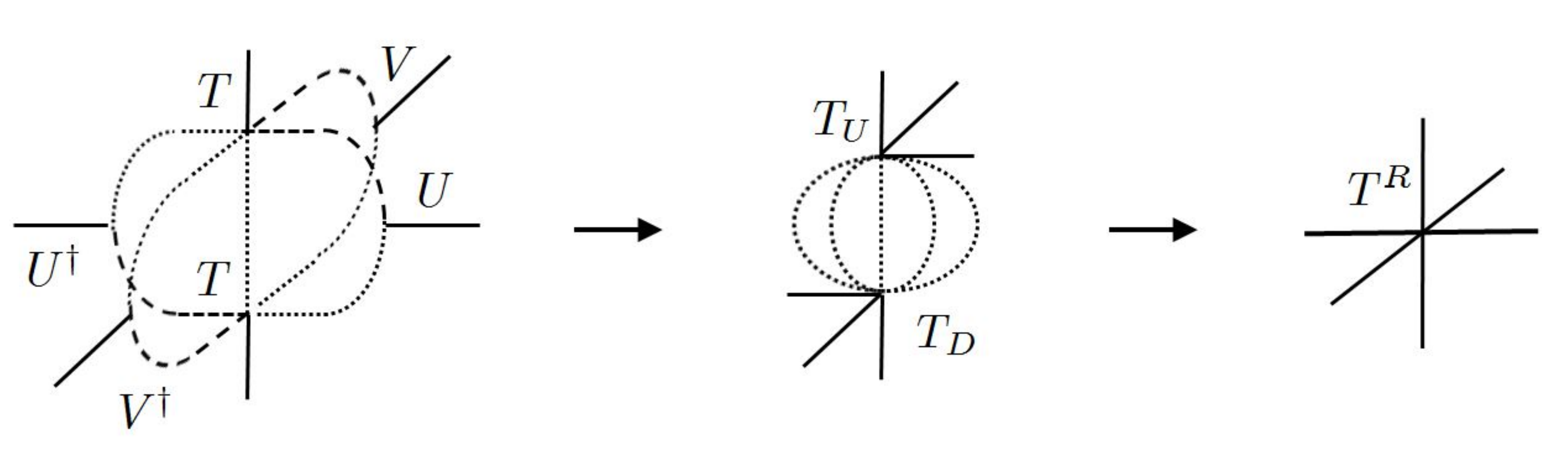}
 \caption{
Final contractions in the HOTRG.
}
\label{fig:cont}
\end{center}
\end{figure}
Figure \ref{fig:cont} shows a procedure of making $T^R$ in three dimensions. 
$T_U$ is a tensor
made of upper $T$ and $U, V$, 
while $T_D$ is one made of lower $T$ and $U^\dag, V^\dag$. 
The cost of making $T_U$  (or $T_D$) is $O(D^{4d-2})$ because
their order is $3d-1$ and inner $d-1$ links are contracted.  
The contraction between $T_U$ and $T_D$ takes $O(D^{4d-1})$
which is the dominant cost of the HOTRG method.

The costs for isometries and final contractions can be reduced with the RSVD 
although systematic errors could be larger.

\section{Extension to higher dimensions}
\label{app:d_Triad_RG}

It is straightforward to extend our method to any $d$ dimension for $d \ge 2$.
We may give a $d$-dimensional triad unit as
\begin{eqnarray}
&& T_{x_1x_1^\prime x_2 x_2^\prime \cdots x_d x_d^\prime } = A^{(1)}_{x_1x_2 a_1} A^{(2)}_{a_1 x_3 a_2}\cdots 
A^{(d-1)}_{a_{d-2} x_d a_{d-1}} \nonumber \\
&& \hspace{1cm}  \times A^{(d)}_{a_{d-1} x_d^\prime a_d} \cdots  A^{(2d-3)}_{a_{2d-4} x^\prime_{3} a_{2d-3}} 
A^{(2d-2)}_{a_{2d-3} x^\prime_2 x^\prime_1}.  
\end{eqnarray}
Then the similar calculation as done in three dimensions tells us that
$K^{(\mu)}$ in the $\mu$ direction ($\mu=1,2\cdots,d-1$), 
which is defined as eq.(\ref{Kmat}), 
 is given by
\begin{eqnarray}
&& K^{(\mu)} = \tilde S_{\mu-1} S_\mu \cdots S_{d-1} R_{d-1} \nonumber \\
&&\hspace{1cm} \times (\tilde S_{\mu-1} S_\mu \cdots S_{d-2} R_{d} )^T,
\label{K_for_d}
\end{eqnarray} 
where
\begin{eqnarray}
(S_n)_{ij,kl} = \sum_{m} A^{(n)}_{imk} (A^{(n)}_{jml})^*,
\end{eqnarray}
and
\begin{eqnarray}
&& \hspace{-5mm}  (\tilde S_n)_{ab,cd}= \sum_{i,j,k} A^{(n)}_{iac} (A^{(n)}_{jbd})^* (\tilde S_{n-1})_{kk,ij}, \\
&& \hspace{-5mm} (R_n)_{ab,cd}= \sum_{i,j,k} A^{(2d-n-1)}_{aci} (A^{(2d-n-1)}_{bdj})^* (R_{n-1})_{ij,kk}, 
\end{eqnarray}
with $(\tilde S_0)_{ab,cd}=(R_0)_{ab,cd}=\delta_{ac}\delta_{bd}$. 
We should note that $\tilde S_n$ and $R_n$ are determined recursively 
at an $O(D^5)$ cost.
The cost of evaluating $K^{(\mu)}$ from the matrix products in 
eq.(\ref{K_for_d}) is $O(D^6)$ and the cost of diagonalizing $K^{(\mu)}$ is also $O(D^6)$. 
We thus find that the power of cost for making isometries does not depend on the dimensionality 
but on $O(D^6)$ in any dimension. 
This cost is reduced to $O(D^5)$ with the RSVD 
as well as the three-dimensional case. 
  
The renormalized tensor $T^R$ for $d$ dimension is given in the 
same manner as eq.(\ref{TR}) with isometries $U^{(\mu)}$. 
We can show that 
\begin{eqnarray}
&& T^R_{x_d x_d^\prime x_1x_1^\prime \ldots x_{d-1} x_{d-1} ^\prime} = \sum_{a,b,c,d} 
{\cal D}_{x_d x_1 \cdots x_{d-1} a b } \nonumber \\
&&\hspace{2cm} \times {\cal M}_{abcd} \, {\cal U}_{cd x_2^\prime \cdots x_{d}^\prime x_1^\prime}, 
\label{TR_triad_for_d}
\end{eqnarray}
where ${\cal M}_{abcd}=\sum_w A^{(d-1)}_{awc}  A^{(d)}_{bwd} $, and $d+2$th order tensor ${\cal U}$ 
is made of $A^{(d)}$, $A^{(d+1)}$,$\cdots$, $A^{(2d-2)}$ and $U^{(\mu)}$ 
such as eq.(\ref{U_def}) while ${\cal D}$ is made of the other tensors. 
Since  ${\cal U}$ and ${\cal D}$ are tensors of order $d+2$, 
the naive memory usage is proportional to $O(D^{d+2})$ if they 
are stored directly on the memory. 
It is easy to show that these tensors are created at $O(D^{d+3})$ costs 
by taking contractions in the appropriate order.
Repeating procedures as shown 
in eqs.(\ref{step1})-(\ref{step3}),  
one can obtain a triad representation of  $d$-dimensional $T^R$ 
at an $O(D^{d+3})$ cost.

\end{document}